\documentclass[aps,pra,amsfonts,superscriptaddress]{revtex4} %,
\usepackage{times}
\usepackage{bbm,amsmath} 
\usepackage{hyperref}

\usepackage[]{graphicx}
\newcommand{\eq}{\begin{eqnarray}} 
\newcommand{\en}{\end{eqnarray}}
\def\bra#1{\mathinner{\langle{#1}|}}
\def\ket#1{\mathinner{|{#1}\rangle}}

\newcommand{\braket}[2]{\langle #1|#2\rangle}

\begin{document}

%\keywords{Entanglement, Indistinguishability, Detection process, Quantum and classical correlations, Symmetrization postulate}

\title{Entanglement of Identical Particles and the Detection Process}

\author{Malte C. Tichy\footnote{Corresponding author\quad E-mail:~\textsf{tichy@phys.au.dk},}}
\address{Physikalisches Institut, Albert--Ludwigs--Universit\"at Freiburg, Hermann--Herder--Str.~3, D--79104 Freiburg, Germany}
\address{Lundbeck Foundation Theoretical Center for Quantum System Research, Department of Physics and Astronomy, University of Aarhus, DK--8000 Aarhus C, Denmark}

\author{Fernando de Melo}
\address{Physikalisches Institut, Albert--Ludwigs--Universit\"at Freiburg, Hermann--Herder--Str.~3, D--79104 Freiburg, Germany}
\address{Instituut voor Theoretische Fysica, Katholieke Universiteit Leuven, Celestijnenlaan 200D, B--3001 Heverlee, Belgium}
\address{Centrum Wiskunde \& Informatica, Science Park 123, NL-1098 XG Amsterdam, Netherlands}

\author{Marek Ku\'s}
\address{Center for Theoretical Physics PAS, Al. Lotnik\'ow 32/46, PL--02-668 Warszawa, Poland}

\author{Florian Mintert}
\address{Physikalisches Institut, Albert--Ludwigs--Universit\"at Freiburg, Hermann--Herder--Str.~3, D--79104 Freiburg, Germany}
\address{Freiburg Institute for Advanced Studies, Albert-Ludwigs-Universit\"at, Albertstrasse 19, D--79104 Freiburg, Germany}
\author{Andreas Buchleitner}
\address{Physikalisches Institut, Albert--Ludwigs--Universit\"at Freiburg, Hermann--Herder--Str.~3, D--79104 Freiburg, Germany}

\begin{abstract}
We introduce {\em detector-level entanglement}, a unified entanglement concept for identical particles that takes into account the possible deletion of many-particle which-way information through the detection process. The concept implies a measure for the  \emph{effective indistinguishability} of the particles, which is controlled by the measurement setup and which quantifies the extent to which the (anti-)symmetrization of the wave-function impacts on physical observables. Initially indistinguishable particles can gain or loose entanglement on their transition to distinguishability, and their quantum statistical behavior depends on their initial entanglement. Our results show that entanglement cannot be attributed to a state of identical particles alone, but that the detection process has to be incorporated in the analysis. 
\end{abstract}

\maketitle               % Produces the title.

\section{Introduction}
Entanglement and the indistinguishability of particles lead to a variety of counter-intuitive quantum many-body phenomena. Entanglement manifests itself most prominently in measurements at remote detectors that exhibit stronger correlations than allowed by local and realistic theories \cite{Bell:1964pt}. Such \emph{quantum correlations} imply the principle impossibility to assign a \emph{physical reality} \cite{Ghirardi:2003uq}, {\it i.e.}~a complete set of physical properties, to an entangled particle individually -- a virtually unsettling consequence \cite{PhysRev.47.777}. For example, the polarization of a photon that is entangled to the electronic state of an atom does not possess a value before its measurement. 

Similarly, it is impossible to assign an identifying label to any particle in a system of two or more identical particles, and the many-body wave-function has to be (anti)symmetrized for (fermions) bosons \cite{Messiah:1964ys}. Again, a postulate of quantum mechanics dictates our very description of reality and sets strict bounds on the information that we can retrieve from a quantum system. In this article, we deal with the \emph{physical} consequences of the symmetrization postulate for entanglement, which are insinuated by the aforementioned analogy. 

For different types of particles, like electrons and protons, {\it i.e.}~non-identical particles, entanglement can be defined rigorously \cite{Werner:1989ve}. Since the particles that carry the entangled degrees of freedom are intrinsically distinct, they have an undetachable, definite \emph{identity} and can thus be \emph{labeled}: they possess well-defined properties such as rest mass or charge by which they can be discriminated unambiguously. Consequently, any measurement on one particle can be assigned to the density matrix that acts on the Hilbert space of this particle. The abstract quantum-information paradigm of parties that control quantum systems \cite{Nielsen:2000fk} -- here: particle properties measured by detectors -- can be applied immediately for an analysis of the physical situation. 

The identification of particles and Hilbert spaces, however, breaks down as soon as identical particles are involved. Other distinctive properties are thus necessary to discriminate the particles. If they are localized far from each other, the particles define two distinguishable entities, related unambiguously to one local detector each. Between these entities, entanglement can then be defined. The (anti)symmetrization procedure does not affect any observable defined on these entities, and can thus be neglected \cite{Herbut:1987hb}. This is already realized in quantum mechanics textbooks, {\it e.g.} \cite{Peres:1993jt} states that \begin{quotation} \emph{``No quantum prediction, referring to an atom located in our laboratory, is affected by the mere presence of similar atoms in remote parts of the universe.''} \end{quotation} 
For example, in a scenario with an atom on the moon that is possibly entangled to an atom on the earth, one can safely neglect the (anti)symmetrization procedure, since the atom on the moon never triggers the detector on the earth, and vice-versa. 

Dealing with the (anti)symmetry of the many-particle wave-function is therefore a rather formal problem when particles are spatially well-separated and clearly distinguishable by this preparation. The apparent correlations in the particle labels must not be accounted for as \emph{physical}, while correlations between remote detectors should. This is achieved by entanglement definitions for identical particles provided in \cite{schliemann-cirac,Eckert:2002vn,Ghirardi-statphys,Ghirardi:2004fk,al:2009cr}, which we will refer to as \emph{a priori} entanglement.

Yet, in other settings, such clear differentiation is sometimes ambiguous or even impossible. Electrons that leave an ionizing helium atom,  photons passing simultaneously through the input arms of a beam splitter in a Hong-Ou-Mandel setting (HOM) \cite{Ou:1988jb}, or atoms in a BEC \cite{Simon:2002ud} are strongly overlapping in space, such that their external states, {\it i.e.}~their spatial preparation,  do not allow to address the particles individually. Their indistinguishability then becomes relevant and affects the results of correlation measurements. In such situations, the impossibility to assign a label to a particle may indeed imply the impossibility to assign a physical reality to a particle that is measured in a detector. The (anti)symmetrization of the wave-function then results in quantum correlations, and cannot be neglected by any means. Quite in general, the dynamical behavior of entanglement during transitions from overlapping, indistinguishable, to well localized, distinguishable particles is heretofore barely understood, while the interest in its role in atomic and molecular \cite{al.:2007df} as well as in biological systems \cite{Cai:2010qf} is growing. Since such natural systems do not conform to a clear paradigmatic setting of remote parties, as in quantum information, it is necessary to include the measurement prescription in the description of entanglement. 

In the present article, we introduce a such unified view on entanglement, which allows the consistent quantification of the entanglement between identical particles in arbitrary settings: Our description comprises the aforementioned situation with separately prepared, thus \emph{distinguishable} particles, but also any possible scenario with particles that strongly overlap in space and that do not suggest  any natural subsystem structure, {\it i.e.}~when they indeed constitute \emph{indistinguishable} particles. In order to quantify the impact of the indistinguishability on the measured quantum correlations, appropriate physical quantities are identified. We show that the indistinguishability of the particles depends on the experimental context, and that it has a manifold impact: Particles can gain or loose quantum correlations through the measurement process.

We first review \emph{a priori} entanglement in Section \ref{sectionapriori} and discuss an experimental situation in which this concept fails. Section \ref{sectiondetectorlevel} then introduces detector-level entanglement and the quantities that govern its behavior. By means of a model-system, we discuss the implications of the indistinguishability of particles for measured correlations  in Section \ref{applications}, and conclude in Section  \ref{conclusions}. 

\section{A priori entanglement} \label{sectionapriori}
In previous approaches to the entanglement of identical particles \cite{schliemann-cirac,Eckert:2002vn,Ghirardi-statphys,al:2009cr}, which we refer to as \emph{a priori} entanglement, the problem of the (anti)symmetrization of the many-particle wave-function is tackled by effectively labeling the particles by some distinctive properties, {\it e.g.~}their external state. In analogy to the Schmidt coefficients, Refs.~\cite{schliemann-cirac,Eckert:2002vn} use the Slater coefficients to determine the entanglement of identical particles. For example, if the minimal number of Slater determinants needed to describe a many-fermion state is not unity, the state is said entangled \cite{al:2009cr}. 

Much like distinguishable particles, two \emph{a priori} non-entangled, \emph{identical} particles will remain always unentangled unless an interaction between them takes place. The interaction between the particles may be direct, or mediated through ancilla particles. While the states of the particles  may change in time, their identity, as given by their initially chosen distinctive properties, will be preserved through unitary time evolution. 

The absence of a priori entanglement implies that it is, \emph{in principle}, possible to perform measurements on a particle in the system by which one obtains a certain known result \emph{deterministically}, {\it i.e.}~one merely \emph{reads off} a pre-existing value of a physical property -- a situation well in accordance with local realism \cite{Ghirardi:2003uq}. A strong and crucial assumption is, however, made here: it is taken for granted that experimentalists are indeed able and willing to always choose those detectors that address the well-defined properties of the particles. 

\begin{figure} \includegraphics[width=\linewidth,angle=0]{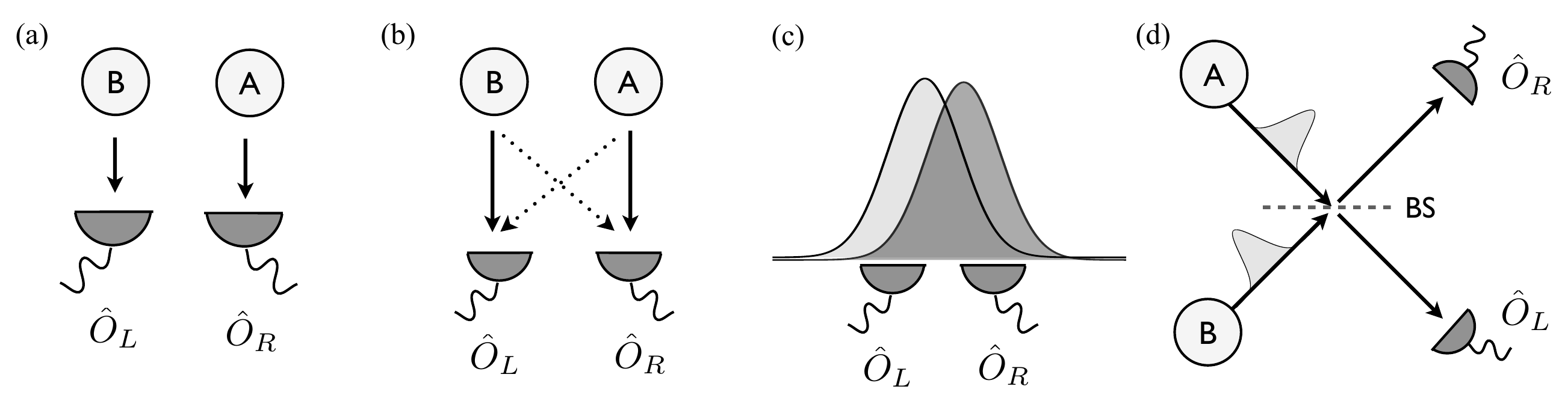}  
\caption{Setups for the measurement of identical particles. (a) Unambiguous scheme, each particle is assigned to exactly one detector. The A (B)-particle will always be detected by the left (right) detector, denoted by $\hat O_L$ ($\hat O_R$). (b) Ambiguous scheme, both particles can trigger either detector. The two two-particle paths are depicted by the solid and the dashed arrows. (c) The generic situation of (b) is realized by two wave-functions that overlap in space, such that either detector can be triggered by either particle. (d) Hong-Ou-Mandel setup, two photons prepared in A and B scatter off the beam splitter (BS) and fall into each output mode with equal amplitude. This setup also realizes the ambiguous scheme shown in (b). } \label{illu} \end{figure}

Particles can be distinguished by their external preparation when they are spatially well separated, such that the assumption is then well justified, since each detector is then assigned exactly one particle. Such situation is also depicted in Figure \ref{illu}(a), where the spatial wave-functions of two particles have a finite overlap with exactly one detector each. The experimentalists then simply read off the preparation of the spin state of the particle in one of the detectors. We will refer to such a situation as an \emph{unambiguous} setup. 

In many situations, like, {\it e.g.}, the one depicted in Figure \ref{illu}(b,c), the external degrees of freedom and the detectors do, however, not coincide anymore, and the \emph{a priori} assumption is no longer justified -- the experimentalists then choose to not merely read off pre-existing, known values, but to use an \emph{ambiguous} detector setting in which the particles appear as indistinguishable. In an HOM-experiment, illustrated in Figure \ref{illu}(d), entanglement in the output arms of a beam splitter is created by the passage of two unentangled photons of opposite polarization through both input arms, and further post-selection on the coincident events \cite{Ou:1988jb,Bose:2002vf}. The initial state of the system reads 
\eq \ket{\Psi_{\mathrm{HOM,ini}}}=\frac{1}{\sqrt 2}\left( \ket{A,H;B,V}+\ket{B,V;A,H} \right), \label{HOMini} \en 
where $\ket{A,H}$ ($\ket{B,V}$) denotes the quantum state of a photon in the input mode $A$ ($B$) with horizontal (vertical) polarization, and $\ket{\Psi;\Phi}$ is the short-hand notation for $\ket{\Psi}_1 \otimes{\ket{\Phi}}_2$.  At first sight, $\ket{\Psi_{\mathrm{HOM,ini}}}$ might seem to be entangled since it cannot be written as a single factor. It is, however, correctly recognized as non-entangled by the a priori definitions \cite{schliemann-cirac,Eckert:2002vn,Ghirardi-statphys}: a measurement of the polarization of the photons at the input arms indeed does not exhibit any correlations, the photon in mode $A(B)$ is always horizontally (vertically) polarized. The correlations in the particle labels in (\ref{HOMini}) -- particle ``1'' seems to be entangled with particle ``2'' -- are not physical, since no physical operator can be implemented to measure the property of a particle that is specified by the mere label. 
 
The time evolution of the external quantum states induced by the scattering on the beam splitter reads, by virtue of Fresnel's equations \cite{Hecht:1987uq}, 
\eq 
\ket{A}& \rightarrow & \frac{1}{\sqrt 2} \left( \ket{L}+\ket{R} \right) =: \ket{\tilde A} \label{LR1} ,\\ 
\ket{B}& \rightarrow & \frac{1}{\sqrt 2} \left( \ket{L}-\ket{R} \right) =: \ket{\tilde B} \label{LR2} ,
\en
where $\ket{L}, \ket{R}$ denote the output modes of the beam splitter. With the above definition of $\ket{\tilde A}$ and $\ket{\tilde B}$, the final state of the photons in the output arms becomes
\eq \ket{\Psi_{\mathrm{HOM,fin}}}=\frac{1}{\sqrt 2}\left( \ket{\tilde A,H;\tilde B,V}+ \ket{\tilde B,V;\tilde A,H}\right) \label{HOMfin} .\en 
The functional form of (\ref{HOMini}) and (\ref{HOMfin}) is apparently unchanged during the scattering, and we can understand (\ref{HOMfin}) with a mere re-labeling of the modes applied on (\ref{HOMini}). This is also immediate from the fact that only one-particle unitary evolutions occur, and no interaction  between the photons takes place: initially orthogonal quantum states of two particles will necessarily remain orthogonal in the absence of mutual interaction. 
Entanglement measures such as in \cite{schliemann-cirac,Eckert:2002vn,Ghirardi-statphys}, which only account for the form of the quantum state, vanish, both for the initial state (\ref{HOMini}) and for the final state (\ref{HOMfin}). The physical situation before and after the scattering on the beam splitter is, however, totally distinct: Both photons have a finite probability to be detected by the \emph{same} local detector \emph{after} the scattering process, whereas \emph{before}, a local detector placed at one local optical mode can only detect one of the particles. 

Since the two two-particle paths ($A\rightarrow R, B\rightarrow L$) and ($A\rightarrow L, B\rightarrow R$) are indistinguishable, they are \emph{coherently} superimposed, and entanglement between the particles is recorded for those detection events where both detectors click \emph{coincidentally} \cite{PhysRevLett.61.2921,Ou:1988jb}. The emerging quantum correlations are also immediate when we consider the state obtained upon post-selection of the coincident events, {\it i.e.}~when one particle is found in each output arm:
\eq \ket{\Psi_{\mathrm{HOM,fin,proj}}}&=&\frac{1}{2} \left( - \ket{L,H; R,V} +\ket{L,V; R,H} -  \ket{R,V; L,H} +\ket{R,H; L,V} \right) .   \label{homfinproj} \en
This form shows that the polarization of the photon found in the state $\ket{L}$ is correlated with the polarization of the one detected in $\ket{R}$. The quantum correlations recorded by the detectors can be explained by the \emph{ambiguity} of the detector setting: Particles are not distinguished by their spatial states $\ket{\tilde A}, \ket{\tilde B}$, but they are measured by the detectors, $\ket{L}\bra{L}$ and $\ket{R}\bra{R}$, with which each photon has a finite overlap.

It is important to retain that the entanglement in the state (\ref{homfinproj}) crucially relies on the symmetrization of the two-photon wave-function, which results in the indistinguishability of the two two-particle paths shown in Figure \ref{illu}(b) and in a \emph{coherent} superposition as in (\ref{homfinproj}). If non-identical particles are initially prepared in both input arms of the beam splitter, one still looses the information on the internal state of the measured particles and finds anti-correlations between the local measurement outcomes. The \emph{nature} of these anti-correlations, however, is then purely classical: The setup corresponds to a classical machine that randomly distributes pairs of particles to the two detectors. An every-day analogy with colored marbles or socks then captures the essence of the process \cite{Bell:1987uq}. 

In view of this experiment and of similar situations in which the spatial states of particles do not have an unambiguous, one-to-one relationship with the detectors, a characterization of entanglement that incorporates the measurement process as described above is necessary to overcome the discrepancies between hitherto available entanglement measures and entanglement that is actually detected in a specific experimental situation. 

\section{Detector-level entanglement} \label{sectiondetectorlevel}
For this very purpose, we introduce \emph{detector-level entanglement}, which incorporates the effects of the measurement process itself, and therefore does not suffer from the problems mentioned above. 
In contrast to earlier treatments \cite{Ghirardi:2003uq}, we do not assign an absolute value of entanglement to a state on its own. Instead, we identify the particles as those entities which are eventually measured by some specified, distinct detectors, which we assume to have 100\% detection efficiency and to be represented by mutually orthogonal projection operators. For an \emph{unambiguous} detector scheme, each detector is associated with exactly one of the particles, which can then be treated as distinguishable such that \emph{detector level entanglement} reduces to the one determined by the previously introduced criteria \cite{Ghirardi-statphys}. In contrast, an \emph{ambiguous} detector setting, in which each particle can trigger each detector with a certain probability, can erase information about the provenience of the particles: the identification of the particles measured by the detectors with the particles prepared in the external states breaks down, since two \emph{distinct} and possibly indistinguishable two-particle paths contribute to the event characterized by the detection of one particle in each detector, as illustrated in Figure \ref{illu}(b).

Our subsequent considerations apply for a general setting of arbitrarily many detectors, which we assume to detect particles in states of their external, {\it i.e.}~spatial, degrees of freedom. The particles additionally carry an internal degree of freedom such as spin, in which they may be entangled. For the sake of clarity, we restrict ourselves hereafter to the case of two particles and two detectors. While the first- and second-quantized formulations of the problem lead to the same physical picture \cite{Tichy:2011fk}, it is here most convenient to use the suggestive language of first quantization in which most mathematical expressions possess a simple, intuitive form.

\subsection{Detector-level density matrix}
Experimentally observed expectation values are associated with the observable 
\eq \hat O_d(\hat \alpha, \hat \beta)
&=& \underbrace{ \hat O_{L} \otimes \hat \alpha}_{1} \otimes \underbrace{ \hat O_{R} \otimes \hat \beta }_{2} +  \underbrace{ \hat O_{R} \otimes \hat \beta }_{1} \otimes \underbrace{ \hat O_{L} \otimes \hat \alpha}_2\ , \label{expop}\en 
where $\hat O_L$ and $\hat O_R$ are two external, mutually orthogonal projectors and describe spatial detectors we call \emph{left} and \emph{right}, respectively. $\hat \alpha$ and $\hat \beta$ are (not necessarily distinct) observables on the internal spaces of the particles. The terms labeled by 1 (2) act on the external and internal spaces of the first (second) particle.

The two detectors assign an identity (left and right) to each of the detected particles, and thereby specify the entities which may or may not carry entanglement. The \emph{detector-level entanglement} of a state $\hat \rho_a$, with respect to a given set of detectors, can thus be inferred by application of any entanglement measure on the detector-level density matrix $\hat \rho_d$, \emph{reconstructed by quantum state tomography} \cite{James:2001zl} \emph{in the local basis defined by the detectors}. This corresponds to the measurement of a complete orthonormal set of observables, $\hat \chi_i, \hat \chi_j$, on the internal degrees of freedom of the particles, with the two external detectors given by (\ref{expop}): 
\eq \hat \rho_d = N \sum_{i,j} \hat \chi_i \otimes \hat \chi_j \mbox{Tr}\left( \hat O_d(\hat \chi_i, \hat \chi_j) \hat \rho_a \right) , \label{tomo} \en
where the normalization $N$ ensures that $\mbox{Tr}(\hat \rho_d)=1$. Thus, (\ref{tomo}) describes the density matrix of the internal degrees of freedom as reconstructed by the detection procedure. In contrast to previous approaches, our formulation takes into account that the spatial overlap of identical particles, and hence their distinguishability, might change with time, even if their quantum state may not. The density matrix $\rho_d$ obtained in (\ref{tomo}) possesses a subsystem structure which always reflects the actual experimental setting. It is thus sensitive to the dynamics of the structure of the state -- and thus to the a priori entanglement of the system --, and to the particle-detector relationship. 

Thereby the problem of the indistinguishability of particles is overcome: Entanglement measures applied on $\hat \rho_d$ determine the \emph{detector-level} entanglement of $\hat \rho_a$ and reproduce the entanglement measured between the two detectors. The encountered quantum correlations may already be present as a priori entanglement in the state, or they may also be induced by the measurement process itself. As we will see below, also an interplay of a priori and measurement-induced entanglement may take place. 

\subsection{Path weights} \label{pathweigths}
In the actual reconstruction (\ref{tomo}) of the detector-level density matrix $\rho_d$, several terms appear, which can be given physical interpretation. 

In the first place, the probability for two particles injected into the (not necessarily orthogonal) external quantum states $\ket A$ and $\ket B$ to trigger a coincident event is governed by the following quantities: 
\eq D_{LR} &\equiv& \bra A \hat O_L \ket A\bra B \hat O_R \ket B,\\ 
D_{RL} &\equiv& \bra A \hat O_R \ket A\bra B \hat O_L \ket B .  \label{dlrdef} \en
$D_{LR}$  $(D_{RL})$ is the probability for the particle prepared in $\ket A$ ($\ket B$) to be detected in the left detector, while the particle prepared in $\ket B$ ($\ket A$) is detected in the right detector. We will refer to these quantities as \emph{path weights}. The two distinct two-particle paths shown in Figure \ref{illu}(b) correspond to these two events. If one path weight vanishes, a coincident event in the detectors reveals full information about which external state the particles were initially prepared in. The scheme is then \emph{unambiguous}, and the situation boils down to the scheme depicted in Figure \ref{illu}(a). 
 If both path weights, $D_{LR}$ and $D_{RL}$, are non-zero, a click in one of the detectors does not completely reveal the provenience of the particles anymore, which corresponds to an ambiguous setting. 

\subsection{Effective indistinguishability}
Even in the ambiguous case, however, particles can still be effectively distinguishable, since one may discriminate the initial preparations by a further measurement: If $\bra B  \hat O_{L/R}^\dagger \hat O_{L/R} \ket A=0$, the scalar product of the projected states $\hat O_{L/R} \ket A$ and $\hat O_{L/R} \ket B$ vanishes, the states are orthogonal after detection. They can thus be distinguished in the sense that their previous preparation could be inferred by another measurement, {\it i.e.}~the which-way information has not been deleted. At the same time, expectation values for measurements on the internal degrees of freedom at the detectors are not affected by the (anti)symmetrization of the wave-function since all cross-terms involving expressions with factors like $\bra{A}\hat O_L\ket{B}\bra{B}\hat O_R \ket{A}$ necessarily vanish. An analogous scenario with distinguishable particles and symmetrized observables would hence yield the very same expectation values. The (anti)symmetrization of the state only plays a role if the overlap of the projected states $\hat O_{R/L} \ket{A}$ and $\hat O_{R/L} \ket{B}$ does not vanish at both detectors. 
A quantitative description for the impact of indistinguishability is therefore given by the product of the overlaps at both detectors: 
 \eq \gamma \equiv \bra A \hat O_L \ket B \bra B \hat O_R \ket A, \label{effindi} \en where we used that $\hat O_{L/R}^\dagger \hat O_{L/R}=\hat O_{L/R}$, for projector-valued $\hat O_{L/R}$. We baptize $\gamma$ the \emph{effective indistinguishability}. It quantifies how strongly the measurement process erases information about the previous preparation of the particles, and thus provides a measure for the indistinguishability of the particles. In other words, the larger $\gamma$ is, the more coherent the two two-particle paths ($A\rightarrow L, B \rightarrow R$) and ($A\rightarrow R, B \rightarrow L$) -- see Fig.~\ref{illu}(b) -- are super-imposed. The effective indistinguishability includes, in contrast to previous approaches \cite{Sun:2009dk} to distinguishability, the dependence on the detector setting. The case $\gamma=0$ corresponds to the situation above in which particles are effectively distinguishable, and the information on the provenience of the particles is preserved during the detection process. For  $\gamma\neq 0$, the (anti)symmetrization of the wave-function does, however, affect expectation values of physical observables. 

We stress that the path weights $D_{LR}$, $D_{RL}$ and the effective indistinguishability $\gamma$ constitute independent parameters. Only the maximal modulus of $\gamma$ depends on the path weights, which follows from the Cauchy-Schwarz inequality: 
\eq |\gamma| &=& |\bra A \hat O_L \ket B \bra B \hat O_R \ket A| \nonumber  \\
 &=&  |\bra A \hat O_L \hat O_L  \ket B \bra B \hat O_R\hat O_R \ket A| \nonumber \\
  &\le & | \hat O_L \ket{A} | \cdot | \hat O_R \ket{B} | \cdot | \hat O_L \ket{B} | \cdot | \hat O_R \ket{A} |  \nonumber \\ &=& \sqrt{D_{LR} D_{RL}} = : \gamma_{\text{max}} \en

The absolute value of $\gamma$ measures how strongly which-way information is erased by a measurement: 
For $|\gamma|=\gamma_{\text{max}}$, the two states $\hat O_{R/L} \ket A, \hat O_{R/L} \ket B$ are linearly dependent.  Hence, one cannot design any discriminating measurement in this case. For particles which are prepared initially in linearly independent external quantum states, such detection erases the information about their provenience. They are hence completely indistinguishable for such a detector setting.

For particles with some spatial overlap at the detectors, situations occur in which $\gamma\neq 0$, even if $\braket{A}{B}=0$.  In other words, the detection setup can effectively inhibit the differentiation of the particles that was possible before the measurement process. In turn, for any non-orthogonal states with $0<|\braket{A}{B}|<1$, there are projectors which \emph{do} differentiate the two states. Physical distinguishability hence strongly depends on the settings of the detectors. For our purpose of exploring the consequences of indistinguishability on entanglement, it is sufficient to consider real and negative $\gamma$, which corresponds to the case of HOM interferometry. 

Physically speaking, the balance of the path weights $D_{LR}$ and $D_{RL}$ governs the ambiguity of the setup, whereas the effective indistinguishability $\gamma$ quantifies the coherence of the two two-particle paths. The (anti)symmetrization of the wave-function has an impact on  observables only if the path weights do both not vanish \emph{and} $\gamma \neq 0$. 

\section{Applications}  \label{applications}
Let us now discuss the implications of the above for two identical particles prepared in external quantum states $\ket A$, $\ket B$. We assume that $\braket{A}{B}=0$ and take the internal degree of freedom of the particles to be equivalent to a spin-$1/2$-system. The following paradigmatic state will turn out to be strongly affected by ambiguous detector settings:
\eq
 \ket{\Psi} = \frac 1 {\sqrt 2}  \left\{ \cos \epsilon \ket{A, \uparrow;B, \downarrow} + \sin \epsilon \ket{A,\downarrow;B, \uparrow}  + \delta 
\left( \cos \epsilon \ket{B, \downarrow;A, \uparrow} + \sin \epsilon \ket{B, \uparrow;A,\downarrow} \right) \right\}\  \label{ourstate}
\en
The parameter $\delta=1(-1)$ refers to bosons (fermions), $\epsilon$ is real and continuous and controls the \emph{a priori} entanglement between the particles. The entanglement measure concurrence \cite{Wootters:1998fk} of (\ref{ourstate}) reads \eq C_a(\epsilon)=2|\cos \epsilon \sin \epsilon| .  \label{concurrenceapriori} \en  
For $\epsilon=0,\pi/2$, the state is a priori non-entangled, and the particles in $\ket{A}$ and $\ket{B}$ can each be assigned a certain definite spin. For $\epsilon=\pi/4$, the state is maximally entangled, no information at all is available on the results of a spin-measurement on either particle.  

\subsection{Exchange interaction}
Let us first investigate the interplay of a priori entanglement and the bosonic/fermionic nature of the particles. For this purpose, we transfer a text-book example for the exchange interaction \cite{Griffiths:1995fu} to the realm of a priori entangled identical particles in the state (\ref{ourstate}). The expectation value of the squared distance of the particles, $(\hat x_1 \otimes \mathbbm{1}_2-\mathbbm{1}_1 \otimes \hat x_2)^2$,  reads 
 \eq \bra \Psi ( 
 \hat x_1 \otimes \mathbbm{1}_{2} -  \mathbbm{1}_{1}  \otimes \hat x_2
 )^2 \ket \Psi& =& \overbrace{\bra A \hat x^2 \ket A +\bra B \hat x^2 \ket B - 2 \bra A \hat x \ket A \bra B \hat x \ket B}^{(\Delta x_{\text{class}})^2} \nonumber \\ && -\delta 4 \cos \epsilon \sin \epsilon |\bra A \hat x \ket B |^2 \label{exchangeint} , \en where the identity $\mathbbm{1}_s$ that acts on the spin degree of freedom is omitted, for readability.
 
The first line equates the intuitive expectation value for the squared distance $(\Delta x_{\text{class}})^2$ for a state $\ket{A; B}$ of distinguishable particles. In addition, we find a correction that makes the particles appear closer together or further apart, depending on the sign of $\delta \cos \epsilon \sin \epsilon$. This term is  similar to the contribution that is found for non-entangled identical bosons and fermions \cite{Griffiths:1995fu}, and it appears only when there is a finite spatial overlap of the particles, such that $\bra{A}\hat x \ket{B}$ does not vanish, hence it matches the common sense interpretation of ``exchange interaction''. 

In contrast to non-entangled bosons or fermions, the exchange term does not depend only on the species $\delta$ (bosonic or fermionic) of the particles, but also on the parameter $\epsilon$ that governs the a priori entanglement.  By suitably manipulating their a priori entanglement properties, we can thus make bosons behave like fermions, and vice-versa.  It is the \emph{interplay} of the (anti-)symmetrization of the wave-function with the a priori entanglement between the particles that leads to this impact on the expectation value of the inter-particle distance. For $\epsilon=0$, the state describes two particles of definite and orthogonal spin, and no exchange interaction takes place -- indeed, this latter situation corresponds to two particles that can be distinguished by their spin.

In our detection setting, we find an analogous effect, and the specific choice of the \emph{a priori} entangled state affects the coincident detection rate $T$. The latter is given by the expectation value of $O_d(\mathbbm{1},\mathbbm{1})$, and reads  \eq T=  D_{LR} +  D_{RL} + 4 \delta \gamma \cos \epsilon \sin \epsilon. \en 
The last term in this sum is proportional both to the \emph{a priori} concurrence $C_a(\epsilon)$, Eq.~(\ref{concurrenceapriori}), and to the \emph{effective indistinguishability} $\gamma$, Eq.~(\ref{effindi}). Through the latter it includes the overlap of the single-particle states at the detectors, and the resulting (anti)bunching effects can be interpreted as a further signature of exchange interaction, similar to the contribution in (\ref{exchangeint}). The case $D_{LR}=D_{RL}=1/4$, $\gamma=-\gamma_{\text{max}}$ results in complete indistinguishability, corresponding to the situation in a HOM interferometer: Two photons in the $\ket{ \Psi^+}$ Bell-state ($\epsilon=\pi/4$) always bunch ($T=0$), while in the $\ket {\Psi^-}$ Bell-state ($\epsilon=3 \pi/4)$ they antibunch ($T=1$)  \cite{Bose:2003kx}.

\subsection{Detector-level state}
By virtue of  (\ref{tomo}), the state (\ref{ourstate})  yields the detector-level density matrix,
\begin{equation} \rho_{d}=
\frac 1{2T}\left( 
\begin{tabular}{cccc} 0 & 0 & 0 & 0 \\
0 & $T + a$ & $b$ & 0 \\
0 & $b $ & $T -a$ &0 \\
0 & 0 & 0 & 0 \\
\end{tabular}
\right) \label{mystate},
\end{equation}
in the detector level basis  
\eq \{ \ket{L \uparrow, R \uparrow}, \ket{L \uparrow, R \downarrow},\ket{L \downarrow, R \uparrow}, \ket{L \downarrow, R \downarrow} \} , \en with \eq a &=&  {(D_{LR}-D_{RL}) \cos 2 \epsilon},  \label{refa} \\
b &=& 2 \delta \gamma 
 + 2(D_{LR}+D_{RL})\sin \epsilon \cos \epsilon. \label{refb}  \en
 For unambiguous detector settings, $D_{LR}=0, D_{RL}=1$, we simply recover the spin properties of the \emph{a priori} state, detecting a particle in the left (right) detector amounts to detecting the particle that was prepared in the external state $\ket{A}$ ($\ket{B}$). In general,  however, the state at detector level, $\rho_d$, is possibly mixed. We stress that this loss of purity is not rooted in the detector inefficiency or any decoherence mechanism, but it is born out by ambiguous detector settings with non-maximal $\gamma$, {\it i.e.}~for which the two two-particle paths are not super-imposed in a fully coherent way. 
 
The quantitative and qualitative consequences of the loss of which-way information in the setup are captured by the following quantities:
\begin{itemize}
\item The \emph{single-particle predictability} $P_d^{(R/L)}$  \cite{Greenberger88} of the particle detected by the right/left detector,
\eq \left( P^{(R/L)}_d \right)^2 & =& \left|\mbox{Tr}\left(\sigma_z \varrho_{R/L} \right) \right|^2 ,\en
where $\varrho_{R/L}$ is the reduced density matrix that describes the particle in the right/left detector,  
quantifies the observer's ability to predict the outcome of a measurement in the $\sigma_z$-basis. For our setup, the coherences of the reduced density matrices $\varrho_{R/L}$ always vanish by construction, since each single particle is never prepared in superpositions of eigenstates of $\sigma_z$, according to (\ref{ourstate}). The product of the coherences can be interpreted as the \emph{visibility} \cite{jakob2007cae} or interference capability, and this quantity thus also vanishes for the state under consideration. Consequently, the predictability quantifies all single-particle information in our case, and its value is equal for the two detected particles, 
\eq   P_d^2 := \left( P^{(R)}_d\right)^2 =\left( P^{(L)}_d \right)^2 =\frac{(D_{LR} -D_{RL})^2 \cos^2 2 \epsilon }{ T^2} . \label{predi}  \en 
When the setup is fully ambiguous, $D_{LR}=D_{RL}$, and no information on the provenience of a detected particle is available, such that the predictability vanishes. Also when a maximally entangled state is prepared ($\cos 2\epsilon =0$), no information on the preparation of any particle is available, with the analogous consequence, $P_d^2=0$.

\item The \emph{two-particle linear entropy}, \eq  S_d =1- \mbox{Tr}(\rho_{d}^2),  \label{twoplinentro} \en  
 quantifies the impurity of the two-particle state. In our case, the pair of particles (\ref{ourstate}) is always fully anti-correlated (there is always exactly one particle in each internal state, $\ket{\uparrow }, \ket{\downarrow }$), such that  measurements exhibit anti-correlations in the $\sigma_z$-basis. The two-particle linear entropy $S_d$ directly constitutes a measure for the \emph{classical correlations} that are induced by the ambiguous detection scheme. Note that only for $|\gamma|\neq \gamma_{\text{max}}$ can $S_d$ assume a non-vanishing value, {\it i.e.}~the only source of classical correlations is the incoherent distribution of the particles among the detectors.  
\item  The \emph{detector-level concurrence} $C_d$ \cite{Wootters:1998fk} measures the strength of the quantum correlations between the two detected particles. For our specific detector-level state (\ref{mystate}), it is proportional to the product of the non-diagonal elements of $\rho_d$, 
\eq C_d^2&=& \left( \frac{b}{T} \right)^2 = \frac{4 \left( \delta \gamma  +  (D_{LR}+D_{RL}) \sin \epsilon \cos  \epsilon \right)^2 }{ T^2 } . \label{detconc} \en 
\end{itemize}
For the state under consideration, the three quantities are complementary \cite{jakob2007cae}, by virtue of \eq  1=C_d^2 + P_d^2 + 2 S_d . \label{complementarity} \en Note that this relation does not hold for general density matrices \cite{jakob2007cae}, but only for the detector-level state $\rho_d$ of the specific form (\ref{mystate}) .  

We can thus understand a loss of predictability $P_d$ as an increase of correlations, whose (quantum or classical) nature is defined by the balance of $C_d^2$ and $S_d$.

If $\gamma=0$, the detector-level concurrence equals the a-priori value ($C_a=C_d$), and ambiguous detector settings with $D_{LR}\neq 0 \neq D_{RL}$ can still manifest themselves, but only as a higher two-particle entropy and lower single-particle predictability with respect to the unambiguous case.  The identical particles behave distinguishably, hence just like particles of a different kind, and \emph{classical correlations} are induced by the ambiguous detector setting. In contrast, non-vanishing values of $\gamma$ change the situation dramatically, and the concurrence $C_d$ will be affected as well.  For illustration, let us discuss two exemplary transitions, and their experimental implementations in HOM-type setups.
\subsection{Vanishing predictability}
First, we consider a scenario with vanishing predictability, $P_d=0$, by setting $ D_{LR}=D_{RL}=1/4$, see (\ref{predi}). The setup is thus maximally ambiguous, which  corresponds to the balanced HOM-type beam-splitter setup shown in Figure \ref{illu}(d). Each particle has the same probability to trigger either detector. Neither party can predict at all any single spin measurement outcome, while the measurement outcomes at the two detectors are perfectly anti-correlated.  

The nature of these correlations depends on the effective indistinguishability $\gamma$ and on the a priori entanglement defined by $\epsilon$. For \emph{a priori} non-entangled ($\epsilon=0$) and effectively distinguishable ($\gamma=0$) particles, the two-particle state (\ref{mystate}) is maximally mixed, since $b=0$ by virtue of (\ref{refb}), and the uncertainty is purely classical, $C_d=0$, $2 S_d=1$, Eqs.~(\ref{twoplinentro}) and (\ref{detconc}). This situation corresponds to a setup in which distinguishable particles are randomly distributed among the two detectors. The two two-particle paths are not super-imposed coherently, and the induced correlations are purely classical. 

For effectively indistinguishable  particles ($\gamma=-\gamma_{\text{max}}$), the two two-particle paths are super-imposed in a fully coherent way. Consequently, the state at detector level $\rho_d$ is pure, the single-particle predictability vanishes due to \emph{detector-level} entanglement, and the detector-level concurrence (\ref{detconc}) attains its maximal value. 

The case $\gamma=-\gamma_{\text{max}}$ is analogous to the situation in ideal HOM interferometry, in which pairs of maximally entangled particles are created \cite{Ou:1988jb,Bose:2003kx}. If, on the other hand, the temporal delay between the photons is too large such that the provenience of the particles could be in principle inferred by a time-resolved measurement (and $\gamma=0$), the particles result not to be entangled. That is, by variation of $\gamma$, the correlations are converted from merely classical correlations into \emph{detector-level} entanglement! 

\begin{figure}
\includegraphics[width=.6\linewidth,angle=0]{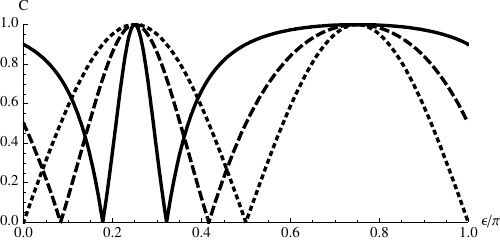} 
\caption{Detector-level concurrence of two bosons in the state (\ref{ourstate}), as a function of $\epsilon$ (which controls the a priori entanglement (\ref{concurrenceapriori}) of the state), for $D_{LR}=D_{RL}=1/4$, $\gamma=0$ (dotted line), $\gamma=-0.5 \gamma_{\text{max}}$ (dashed line), $\gamma=-0.9 \gamma_{\text{max}}$ (solid line). For $\gamma=0$, the detector-level concurrence $C_d$, Eq.~(\ref{detconc}) equates the \emph{a priori} concurrence (\ref{concurrenceapriori}), whereas for $\gamma=-\gamma_{\text{max}}$, $C_d=1$ for all values of $\epsilon$. The transition between distinguishable ($\gamma=0$) and indistinguishable ($\gamma \neq 0$) particles is, however, not always monotonic. }\label{3DCHSH}
\end{figure}

For an a priori non-entangled state with $\epsilon=0$ or $\epsilon=\frac{\pi}{2}$, the detector-level concurrence increases monotonically with $|\gamma|$. The less the particles can effectively be distinguished, the more they are entangled. In contrast, the dependence is non-monotonic and therefore more intricate  for \emph{a priori} entangled states, as illustrated in Figure \ref{3DCHSH}. This behavior is due to the competition between the (measurement-induced) term proportional to $\gamma$, and that proportional to $\cos \epsilon \sin \epsilon$, {\it i.e.}~the a priori concurrence, in the parameter $b$ (\ref{refb}). 

\subsection{Maximal indistinguishability}
As a second example, we now fix the \emph{effective indistinguishability} to $\gamma=-\gamma_{\text{max}}$, and tune the detector bias from unambiguous to completely ambiguous, by variation of $D_{LR}$ between 0 and $1/4$, while  $D_{RL}=(1-\sqrt{D_{LR}})^2$. This transition can be implemented experimentally with several beam splitters that assume different transmission probabilities. Increasing $D_{LR}$ corresponds to smoothly breaking the unambiguous bond between particle and detector, while always maintaining maximal indistinguishability. The induced uncertainty for single spin-measurements is then maximal for $D_{LR}=D_{RL}=1/4$. 

\begin{figure}
\includegraphics[width=.6\linewidth,angle=0]{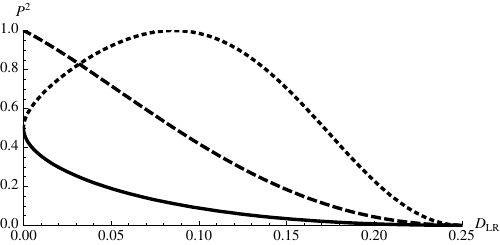} 
\caption{Squared detector-level single particle predictability $P_d^2$ as a function of $D_{LR}$, for $\gamma=-\gamma_{\text{max}}$, $\epsilon=3\pi/8$ (dotted line), $\epsilon=0$ (dashed line) and $\epsilon=5 \pi/8$ (solid line). The predictability is not always a monotonically decreasing function of $D_{LR}$ ($D_{LR}=1/4$ corresponds to a completely ambiguous detector setting). 
} \label{predictability} \label{predicti}
\end{figure}

 No classical uncertainty is created by the detection, since $\gamma=-\gamma_{\text{max}}$, and $\rho_d$ remains pure. In other words, the two two-particle paths are always coherently super-imposed, whereas their amplitudes are varied by $D_{LR}$ and $D_{RL}$. If $D_{LR} \neq D_{RL}$, the occurrence of a coincident event itself contains some statistical information on the possible provenience of the particles, the setup is then not maximally ambiguous and the \emph{detector-level} entanglement is not maximal. While the measurement destroys the information about the preparation of the detected particles totally ($\gamma=-\gamma_{max}$), the probability for a certain type ($\ket A$ measured in the left or right detector) of coincident event  to occur is biased, {\it i.e.}~the two possible two-particle paths shown in Figure \ref{illu}(b) have different amplitudes. Again, predictability (and, due to the above-mentioned complementarity, concurrence) does not depend on $D_{LR}$ monotonically, with remarkable implications: when bunching occurs, \emph{i.e.} $\epsilon< \pi $, the detector-level concurrence $C_d$ can be smaller than the \emph{a priori} concurrence $C_a$, and the predictability (\ref{predi}) consequently grows. In other words, an ambiguous spatial measurement that does not discriminate spins can actually \emph{enhance} the probability of a certain spin-measurement outcome with respect to the unambiguous case. We illustrate this behavior in Figure \ref{predicti}, where the squared detector-level predictability is plotted for different values of $\epsilon$, as a function of $D_{LR}$.

Although we focused on bosons throughout this section, the same situation can be recovered for fermions, merely exchanging the sign of the effective indistinguishability,  $\gamma \rightarrow - \gamma$. In other words, the \emph{species} of the particles (bosonic or fermions) merely boils down to the sign of a parameter, and it is the indistinguishability which is mainly responsible for the encountered coherent many-particle effects \cite{Tichy:2011fk}. 

\section{Conclusions} \label{conclusions}
Entanglement and the indistinguishability of particles \emph{interfere} on several levels -- in the literal and in the figurative sense of the word: The very notion of entanglement needs to be adjusted in order to allow for the impossibility to label identical particles and for the consequent loss of subsystem structure. Thus, discriminating degrees of freedom need to be incorporated explicitly. This leads to the concept of \emph{a priori} entanglement, which deals with the assignment of the notion of entanglement to a state on its own. On the other hand, the indistinguishability of identical particles also comes with a \emph{many-body coherence property} which drastically influences the observable entanglement in many situations. These physical consequences of the symmetrization postulate were so far not captured by the available entanglement concepts in the literature. 

Our here introduced notion of \emph{detector-level entanglement} takes into account both aspects, such that it is suitable for the characterization of entanglement in systems in which particles may overlap in space (and where this overlap may be time-dependent under some nontrivial dynamics) and the assumptions for a priori entanglement are not met, since each detector can be triggered by both particles. 
 We introduced with (\ref{tomo}) the density matrix of a quantum state of two identical particles at the level of the detectors, such that in all situations, including the HOM setup described at the outset, the \emph{physical} impact of indistinguishability is taken into account.  Our examples show the important discrepancy between \emph{a priori} entanglement and \emph{detector-level} entanglement, and illustrate how the parameter $\gamma$ provides a measure for the erasure of information on the particle preparation through the detection process. For vanishing $\gamma$, particles can be treated as distinguishable, since the (anti)symmetrization does not affect non-classical correlations. However, for $\gamma\neq 0$ the (anti)symmetrization is crucial for physical observables in general, and, in particular, it affects the entanglement between the measured particles.  
Our quantitative definition of \emph{detector-level} entanglement incorporates these effects and thus matches the experimental reality.

Whereas the indistinguishability of particles is routinely exploited in linear-optics experiments with photons \cite{Pan:2011fk}, it is usually neglected in experiments with interacting constituents, in which the mutual interaction constitutes the source of quantum correlations \cite{Cirac:1995fk}. In the future, the combined action of these effects -- interaction and indistinguishability -- may be probed, {\it e.g.}~in experiments with cold atoms in optical lattices, whose indistinguishability \emph{and} interaction are under control \cite{Sherson:2010fk}: The internal degrees of freedom provide tunable distinguishability, the interaction can be tuned via Feshbach resonances \cite{Leggett:2001mi}.

The understanding and the differentiation of the aforementioned effects is especially important for the characterization of entanglement in systems that do not follow the paradigm of quantum information, such as atoms and molecules \cite{Schoffler:2008cr,al.:2007df,0953-4075-44-19-192001}. In these systems, the encountered correlations are typically rooted in, both, interaction-induced a priori entanglement, and measurement-induced entanglement.

\subsection*{Acknowledgments}
M. C. T. acknowledges financial support by the German National Academic Foundation. F. de M. acknowledges support from the Alexander von Humboldt-Stiftung, the Belgian Interuniversity Attraction Poles Programme P6/02, and Vidi grant 639.072.803 from the Netherlands Organization for Scientific Research (NWO). Partial support of M. K. by Polish NCN grant DEC-2011/01/M/ST2/0379  is acknowledged, as well as support of F. M. by DFG grant MI 1345/2-1. A.B. acknowledges partial support through the EU-COST Action MP1006 "Fundamental Problems in Quantum Physics".

\providecommand{\WileyBibTextsc}{}
\let\textsc\WileyBibTextsc
\providecommand{\othercit}{}
\providecommand{\jr}[1]{#1}
\providecommand{\etal}{~et~al.}


\begin{thebibliography}{[10]}

\bibitem{Bell:1964pt}% article
 \textsc{J.~Bell}\iffalse On the {E}instein {P}odolsky {R}osen paradox\fi,
 \jr{Physics} \textbf{1}, 195 (1964).


\bibitem{Ghirardi:2003uq}% article
 \textsc{G.~Ghirardi} and  \textsc{L.~Marinatto}\iffalse Entanglement and
  properties\fi,
 \jr{Fortschr. Phys.} \textbf{51}, 379 (2003).


\bibitem{PhysRev.47.777}% article
 \textsc{A.~Einstein},  \textsc{B.~Podolsky},  and  \textsc{N.~Rosen}\iffalse
  Can quantum-mechanical description of physical reality be considered
  complete?\fi,
 \jr{Phys. Rev.} \textbf{47}, 777 (1935).


\bibitem{Messiah:1964ys}% article
 \textsc{A.\,M.\,L. Messiah} and  \textsc{O.\,W. Greenberg}\iffalse
  Symmetrization postulate and its experimental foundation\fi,
 \jr{Phys. Rev.} \textbf{136}, B248 (1964).


\bibitem{Werner:1989ve}% article
 \textsc{R.\,F. Werner}\iffalse Quantum states with einstein-podolsky-rosen
  correlations admitting a hidden-variable model\fi,
 \jr{Phys. Rev. A} \textbf{40}, 4277 (1989).


\othercit
\bibitem{Nielsen:2000fk}% book
 \textsc{M.\,A. Nielsen} and  \textsc{I.\,L. Chuang},
Quantum computation and quantum information (Cambridge University Press,
  Cambridge, 2000).


\bibitem{Herbut:1987hb}% article
 \textsc{F.~Herbut} and  \textsc{M.~Vuji{\v c}i\'c}\iffalse Irrelevance of the
  pauli principle in distant correlations between identical fermions\fi,
 \jr{J. Phys. A} \textbf{20}, 5555 (1987).


\othercit
\bibitem{Peres:1993jt}% book
 \textsc{A.~Peres},
Quantum theory: concepts and methods (Kluwer Academic Publishers, New York,
  1993).


\bibitem{schliemann-cirac}% article
 \textsc{J.~Schliemann},  \textsc{J.~Cirac},  \textsc{M.~Ku\'{s}},
  \textsc{M.~Lewenstein},  and  \textsc{D.~Loss}\iffalse Quantum correlations
  in two-fermion systems\fi,
 \jr{Phys. Rev. A} \textbf{64}, 022303 (2001).


\bibitem{Eckert:2002vn}% article
 \textsc{K.~Eckert},  \textsc{J.~Schliemann},  \textsc{D.~Bru\ss},  and
  \textsc{M.~Lewenstein}\iffalse Quantum {C}orrelations in {S}ystems of
  {I}ndistinguishable {P}articles\fi,
 \jr{Ann. Phys.} \textbf{299}, 88 (2002).


\bibitem{Ghirardi-statphys}% article
 \textsc{G.~Ghirardi},  \textsc{L.~Marinatto},  and  \textsc{T.~Weber}\iffalse
  Entanglement and {P}roperties of {C}omposite {Q}uantum {S}ystems: {A}
  {C}onceptual and {M}athematical {A}nalysis\fi,
 \jr{J. Stat. Phys} \textbf{108}, 49 (2002).


\bibitem{Ghirardi:2004fk}% article
 \textsc{G.~Ghirardi} and  \textsc{L.~Marinatto}\iffalse General criterion for
  the entanglement of two indistinguishable particles\fi,
 \jr{Phys. Rev. A} \textbf{70}, 012109 (2004).


\bibitem{al:2009cr}% article
 \textsc{A.\,R. Plastino},  \textsc{D.~Manzano},  and  \textsc{J.\,S.
  Dehesa}\iffalse {Separability criteria and entanglement measures for pure
  states of N identical fermions}\fi,
 \jr{Europhys. Lett.} \textbf{86}, 20005 (2009).


\bibitem{Ou:1988jb}% article
 \textsc{Z.\,Y. Ou} and  \textsc{L.~Mandel}\iffalse Violation of {B}ell's
  {I}nequality and {C}lassical {P}robability in a {T}wo-{P}hoton {C}orrelation
  {E}xperiment\fi,
 \jr{Phys. Rev. Lett.} \textbf{61}, 50 (1988).


\bibitem{Simon:2002ud}% article
 \textsc{C.~Simon},
 \jr{Phys. Rev. A} \textbf{66}, 052323 (2002).


\bibitem{al.:2007df}% article
 \textsc{D.~Akoury},  \textsc{K.~Kreidi},  \textsc{T.~Jahnke},
  \textsc{T.~Weber},  \textsc{A.~Staudte},  \textsc{M.~Sch{\"o}ffler},
  \textsc{N.~Neumann},  \textsc{J.~Titze},  \textsc{L.\,P.\,H. Schmidt},
  \textsc{A.~Czasch},  \textsc{O.~Jagutzki},  and  \textsc{R.\,A. {Costa
  Fraga}}\iffalse The {S}implest {D}ouble {S}lit: {I}nterference and
  {E}ntanglement in {D}ouble {P}hotoionization of {H}$_2$\fi,
 \jr{Science} \textbf{318}, 949 (2007).


\bibitem{Cai:2010qf}% article
 \textsc{J.~Cai},  \textsc{S.~Popescu},  and  \textsc{H.\,J. Briegel}\iffalse
  Dynamic entanglement in oscillating molecules and potential biological
  implications\fi,
 \jr{Phys. Rev. E} \textbf{82}, 021921 (2010).


\bibitem{Bose:2002vf}% article
 \textsc{S.~Bose} and  \textsc{D.~Home}\iffalse Generic {E}ntanglement
  {G}eneration, {Q}uantum {S}tatistics and {C}omplementarity\fi,
 \jr{Phys. Rev. Lett.} \textbf{88}, 050401 (2002).


\othercit
\bibitem{Hecht:1987uq}% book
 \textsc{E.~Hecht} and  \textsc{A.~Zajac},
Optics (Addison-Wesley, Reading, 1987).


\bibitem{PhysRevLett.61.2921}% article
 \textsc{Y.\,H. Shih} and  \textsc{C.\,O. Alley}\iffalse New type of
  einstein-podolsky-rosen-bohm experiment using pairs of light quanta produced
  by optical parametric down conversion\fi,
 \jr{Phys. Rev. Lett.} \textbf{61}, 2921 (1988).


\othercit
\bibitem{Bell:1987uq}% book
 \textsc{J.\,S. Bell},
Speakable and Unspeakable in Quantum Mechanics (Cambridge University Press,
  Cambridge, 1987).


\othercit
\bibitem{Tichy:2011fk}% phdthesis
 \textsc{M.\,C. Tichy},
Entanglement and Interference of Identical Particles, 
PhD thesis, Universit\"at Freiburg,  2011. http://www.freidok.uni-freiburg.de/volltexte/8233/.


\bibitem{James:2001zl}% article
 \textsc{D.\,F. James},  \textsc{P.\,G. Kwiat},  \textsc{W.\,J. Munro},  and
  \textsc{A.\,G. White}\iffalse Measurement of qubits\fi,
 \jr{Phys. Rev. A} \textbf{64}, 052312 (2001).


\bibitem{Sun:2009dk}% article
 \textsc{F.\,W. Sun} and  \textsc{C.\,W. Wong}\iffalse Indistinguishability of
  independent single photons\fi,
 \jr{Phys. Rev. A} \textbf{79}, 013824 (2009).


\bibitem{Wootters:1998fk}% article
 \textsc{W.\,K. Wootters}\iffalse Entanglement of formation of an arbitrary
  state of two qubits\fi,
 \jr{Phys. Rev. Lett.} \textbf{80}, 2245 (1998).


\othercit
\bibitem{Griffiths:1995fu}% book
 \textsc{D.\,J. Griffiths},
Introduction to Quantum Mechanics (Prentice Hall, Upper Saddle River, 1995).


\bibitem{Bose:2003kx}% article
 \textsc{S.~Bose},  \textsc{A.~Ekert},  \textsc{Y.~Omar},
  \textsc{N.~Paunkovi\'c},  and  \textsc{V.~Vedral}\iffalse Optimal state
  discrimination using particle statistics\fi,
 \jr{Phys. Rev. A} \textbf{68}, 052309 (2003).


\bibitem{Greenberger88}% article
 \textsc{D.\,M. Greenberger} and  \textsc{A.~Yasin}\iffalse Simultaneous wave
  and particle knowledge in a neutron interferometer\fi,
 \jr{Phys. Lett. A} \textbf{128}, 391 (1988).


\bibitem{jakob2007cae}% article
 \textsc{M.~Jakob} and  \textsc{J.~Bergou}\iffalse Complementarity and
  entanglement in bipartite qudit systems\fi,
 \jr{Phys. Rev. A} \textbf{76}, 52107 (2007).


\othercit
\bibitem{Pan:2011fk}% unpublished
 \textsc{J.\,W. Pan},  \textsc{Z.\,B. Chen},  \textsc{C.\,Y. Lu},
  \textsc{H.~Weinfurter},  \textsc{A.~Zeilinger},  and  \textsc{M.~Zukowski},
Multi-photon entanglement and interferometry,
arXiv:0805.2853, to appear in {\it Rev. Mod. Phys.}, 2008.


\bibitem{Cirac:1995fk}% article
 \textsc{J.\,I. Cirac} and  \textsc{P.~Zoller}\iffalse Quantum computations
  with cold trapped ions\fi,
 \jr{Phys. Rev. Lett.} \textbf{74}(05), 4091 (1995).


\bibitem{Sherson:2010fk}% article
 \textsc{J.\,F. Sherson},  \textsc{C.~Weitenberg},  \textsc{M.~Endres},
  \textsc{M.~Cheneau},  \textsc{I.~Bloch},  and  \textsc{S.~Kuhr}\iffalse
  Single-atom-resolved fluorescence imaging of an atomic mott insulator\fi,
 \jr{Nature} \textbf{467}, 68 (2010).


\bibitem{Leggett:2001mi}% article
 \textsc{A.\,J. Leggett}\iffalse Bose-einstein condensation in the alkali
  gases: Some fundamental concepts\fi,
 \jr{Rev. Mod. Phys.} \textbf{73}, 307 (2001).


\bibitem{Schoffler:2008cr}% article
 \textsc{M.\,S. Sch\"offler},  \textsc{J.~Titze},  \textsc{N.~Petridis},
  \textsc{T.~Jahnke},  \textsc{K.~Cole},  \textsc{L.\,P.\,H. Schmidt},
  \textsc{A.~Czasch},  \textsc{D.~Akoury},  \textsc{O.~Jagutzki},
  \textsc{J.\,B. Williams},  \textsc{N.\,A. Cherepkov},  \textsc{S.\,K.
  Semenov},  \textsc{C.\,W. McCurdy},  \textsc{T.\,N. Rescigno},
  \textsc{C.\,L. Cocke},  \textsc{T.~Osipov},  \textsc{S.~Lee},  \textsc{M.\,H.
  Prior},  \textsc{A.~Belkacem},  \textsc{A.\,L. Landers},
  \textsc{H.~Schmidt-B\"ocking},  \textsc{T.~Weber},  and
  \textsc{R.~D\"orner}\iffalse Ultrafast probing of core hole localization in
  n2\fi,
 \jr{Science} \textbf{320}, 920 (2008).


\bibitem{0953-4075-44-19-192001}% article
 \textsc{M.\,C. Tichy},  \textsc{F.~Mintert},  and
  \textsc{A.~Buchleitner}\iffalse Essential entanglement for atomic and
  molecular physics\fi,
 \jr{J. Phys. B} \textbf{44}, 192001 (2011).



\end{thebibliography}
\end{document}